\documentclass[reprint,amsmath,amssymb,aps]{revtex4-1}
\usepackage{graphicx}
\usepackage{float}
\usepackage{dcolumn}
\usepackage{bm}
\usepackage[utf8]{inputenc}
\usepackage{subfigure}
\usepackage[table]{xcolor}
\usepackage[export]{adjustbox}[2011/08/13]

\begin{document}
	
	\title{Shortest Paths in Complex Networks: Structure and Optimization}
	\author{G. S. Domingues$^1$}
    \author{C. H. Comin$^2$}
    \author{L. da F. Costa$^1$}
    \affiliation{$^1$S\~ao Carlos Institute of Physics, University of S\~ao Paulo, S\~ao Carlos, SP, Brazil}
    \affiliation{$^2$Department of Computer Science, Federal University of S\~ao Carlos, S\~ao Carlos, SP, Brazil}
    
	\date{\today}
	
	\begin{abstract}
	Among the several topological properties of complex networks, the shortest path represents a particularly important characteristic because of its potential impact not only on other topological properties, but mainly for its influence on several dynamical processes taking place on the network.  In addition, several practical situations, such as transit in cities, can benefit by modifying a network so as to reduce the respective shortest paths.  In the present work, we addressed the problem of trying to reduce the average shortest path of several theoretical and real-world complex networks by adding a given number of links according to different strategies. More specifically, we considered: placing new links between nodes with relatively low and high degrees; to enhance the degree regularity of the network; preferential attachment according to the degree; linking nodes with relatively low and high betweenness centrality; and linking nodes with relatively low/low, low/high, and high/high accessibilities.  Several interesting results have been obtained, including the identification of the accessibility-based strategies as providing the largest reduction of the average shortest path length.  Another interesting finding is that, for several types of networks, the degree-based methods tend to provide improvements comparable to those obtained by using the much more computationally expensive betweenness centrality measurement.
 	\end{abstract}
	
	\maketitle

\section{Introduction}
Network science is a growing and important research area, employed to characterize and model a wide range of real complex systems~\cite{ALBERT}, such as financial market~\cite{kaue2012}; electric power transmission systems~\cite{pagani2013}; interaction between proteins~\cite{ciafre2013}; communication and telephony networks~\cite{onnela2007}; city streets systems~\cite{STRANO2013, domingues2018}; citations in scientific papers~\cite{amancio2012,golosovsky2017} or even links between web pages~\cite{COSTA2011}.
	
In order to better understand these systems, various ways of studying their topology have been suggested. One particularly important measurement is the average shortest path length (ASPL)~\cite{ShortestPath}, that corresponds to the average of the minimal distances between all possible pairs of nodes inside a network. In many real world systems, such as in city streets or power grid networks, to know the shortest path between two elements is crucial to adopt strategies to minimize costs and time of agents moving inside these systems. Thus, it is also interesting to find ways to diminish the average shortest path of a network as a means to improve performance of several types of activities.

In this work we aim at studying how different strategies can be applied to optimize the average shortest path of known theoretical complex network models, such as Barabási-Albert (BA), Erd\H{o}s-Rényi (ER), Watts-Strogatz (WS) and Waxman (WAX), as well as the world-wide airport network.  We consider four different strategies: (i) increasing the degree regularity; (ii) adding links preferentially to node degree; (iii) linking pairs of nodes with low and high betweenness; and (iv) adjusting the accessibility by implementing connections between central and peripheral nodes. It is shown that, among the considered strategies, connecting central and peripheral nodes leads to the fastest decrease in average shortest path length. Also, these approaches have distinct effectiveness depending on the network models, with the BA model resulting in the smallest decrease in ASPL.
	
This work begins by presenting the adopted network models, the concepts of average shortest path and the considered optimization strategies. Next, the results are presented and discussed.
	
\section{Methodology}
    
\subsection{Complex Networks Models}
    
Four known theoretical complex networks models were used in the current study:
\begin{enumerate}
    \item The Erd\H{o}s-Rényi model (ER), in which pairs of nodes are connected with a fixed probability $p$. In this work, $p=0.006$ was used;
    \item The Barabási-Albert model (BA)~\cite{ALBERT}, consisting of a random graph in which the nodes are added one-by-one. Each edge of the newly added node connects to the existing ones with a probability proportional to their degree;
    \item The Watts-Strogatz model (WS), which is generated considering an initial regular ring lattice. Each edge of the lattice is rewired with probability $p$, and is placed between two nodes chosen randomly with uniform probability. We adopt $p=0.4$ in the experiments;
    \item The Waxman model (WAX)~\cite{waxman1988}, yielding geometric networks, constructed by randomly placing nodes on a given space and connecting pairs of nodes with probability $p_{ij}$, given as:
        \begin{equation}
        p_{ij} = \alpha \exp^{-d_{ij}/\beta}
        \end{equation}
        
    where $d_{ij}$ is the distance between nodes $i$ and $j$, and $\alpha$ and $\beta$ are parameters. $\alpha= 0.014$ and $\beta = 0.2$ were used in the experiments.
\end{enumerate}

Henceforth, all networks have $N=1000$ nodes and average degree fixed around $6$.

\subsection{Average Shortest Path}
    
The minimum distance between two nodes A and B in a network, $d_{AB}$, corresponds to the minimum number of edges that need to be traveled when moving from A to B. The average shortest path $L$~\cite{ShortestPath} of a given network is defined as the mean of the shortest distances between all possible pair of nodes.  

\begin{figure}[!htbp]
	\includegraphics[width=\columnwidth]{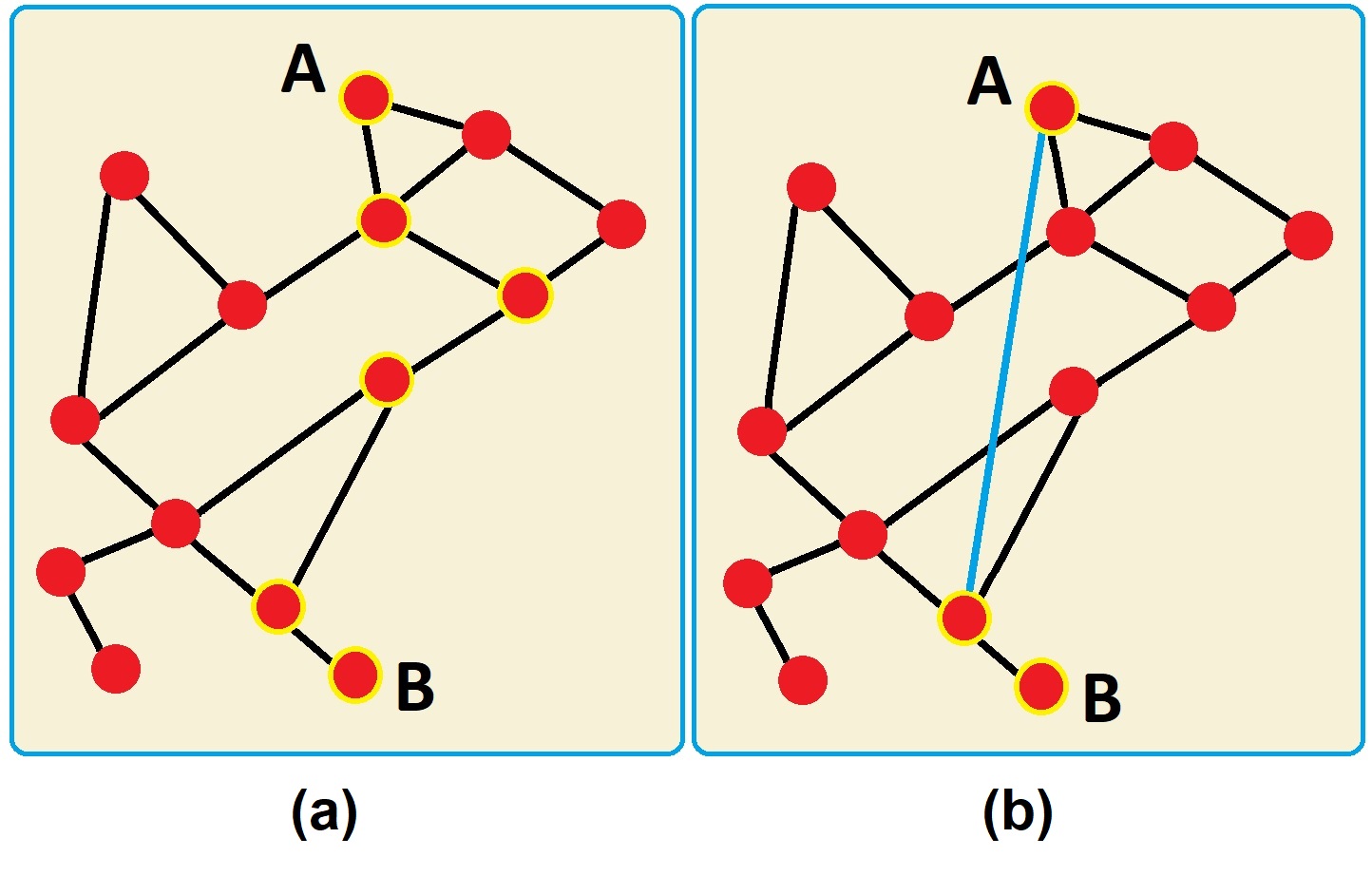}
	\caption{Example of a graph (a) with one of the Shortest Paths between nodes $A$ and $B$ highlighted, evaluated as $d_{AB} = 5$. With addition of a single edge (b, in blue), the Shortest Path is  reduced to $d_{AB} = 2$.}
	\label{LeastPath}
\end{figure}

The shortest path between two nodes can vary substantially as a consequence of small alterations in the network topology, as illustrated in Figure~\ref{LeastPath}.

The average shortest path can be used as an indicator of the effectiveness of some networks with respect to some action, and also as a means for optimizing the respective efficiency (e.g.~\cite{ahuja1995}).  For instance, the facility of pedestrians in moving along a city can be thought of as being inversely related to the average shortest path of the respective complex network representation of the city.

\subsection{Optimization Strategies}

The considered approaches for trying to optimize given complex networks so as to reduce their average shortest path are described as follows.  All considered strategies involve adding a same number of new links $N_a = 50$ to be added to the network.

\textbf{Connecting nodes by degree: } The first strategy used to optimize the average shortest paths of a network involves adding links between nodes with relatively small and large degree, as an attempt to provide alternative, possibly shorter, routes to the paths going through the small degree node.  

\textbf{Making the topology more regular: } The second strategy will be to try to make the degrees of the network more uniform, yielding a new version of the original network which is closer to a regular graph (all nodes having the same degree). In this work, we perform alterations in the original network so as to reduce the node degree standard deviation.  The motivation for this strategy is that regular networks tend to have smaller average shortest path~\cite{buhl2006}. The adopted strategy consists in linking nodes with degree smaller than the current average degree $\left<k\right>$ to a node belonging to its second-neighborhood~\cite{COSTA2006} and also having degree smaller than $\left<k\right>$. 

\textbf{Preferential Attachment:}  Another strategy is to perform a preferential attachment~\cite{cooper2014}, connecting randomly chosen nodes to others following a probability $p_i$ given as $p_i \propto k(i)$, where $k(i)$ is the degree of a node $i$.  The preferential attachment strategy has been motivated by the fact that many real-world networks have a power-law degree distribution that can be modeled using the preferential attachment concept. 
    
\textbf{Betweenness-based approach: } Measurements of centrality~\cite{BRANDES2001} (e.g. Betweenness, Closeness or Stress Centrality) are important for identifying nodes which can be potentially critical, e.g.~in the sense of carrying more information and being more visited by traveling agents. More central nodes also participate in a greater number of shortest paths in comparison with less central nodes.  Thus, one strategy that can be tried for enhancing centrality is to connect nodes having the smallest and largest betweenness centralities.
    
\textbf{Accessibility-based approach: } Given a node $V$, its accessibility $A_h(V)$~\cite{travenccolo2008accessibility} is a measurement of how easily reachable, or accessible, are other nodes at distance $h$ from $V$. This measurement considers the transition probabilities $P_h(V,j)$ of an unbiased random walker going from $V$ to each destination node $j$ at a distance $h$, and is defined by the exponential of the entropy of these probabilities, i.e.:

    \begin{equation}
    A_h(V) = exp(E_h(V,\Omega))
    \end{equation}
where $\Omega$ is the set of all nodes reachable from $V$ in $h$ steps, $N$ the size of set $\Omega$ and $E_h(V,\Omega)$ is given as:
    
    \begin{equation}
    E_h(V,\Omega) = -\sum_{j=1}^{N} P_h(V,j)\log (P_h(V,j))
    \end{equation}

\begin{figure*}[!htbp]
	\includegraphics[width=\linewidth]{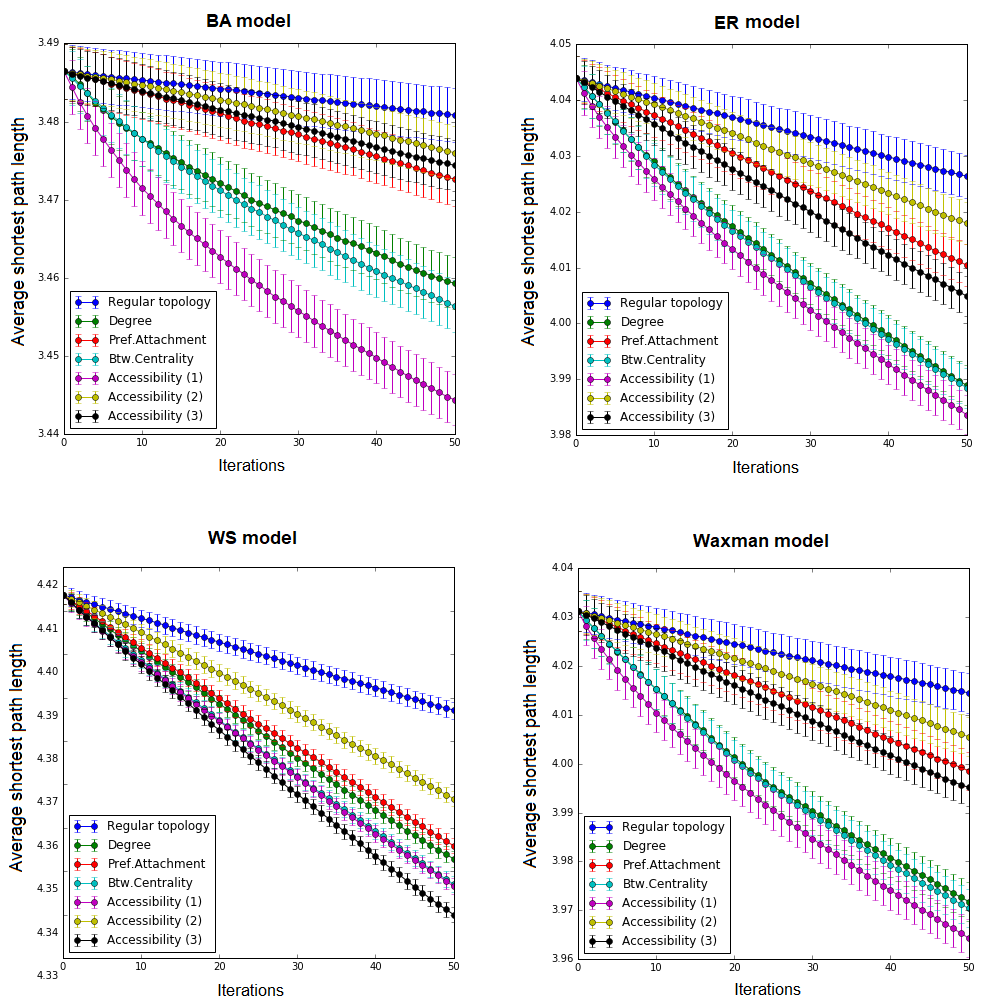}
	\caption{Average shortest path length in terms of the number of added connections (iterations) for each of the considered networks topologies. The error bars are shown as vertical lines to $10\%$ of their real length.}
		\label{resultadoComparacao}
\end{figure*}

The accessibility can be used to identify borders on networks. More specifically, nodes in the more central region of the network tend to present higher accessibility values, while nodes belonging to the periphery/border tend to have smaller accessibility.~\cite{travenccolo2008accessibility}.   This suggests that we can try to enhance the accessibility of the nodes in a network by considering the following strategies: (a) adding connections between central and peripheral nodes - which can be called \emph{accessibility (1)}, for simplicity; (b) interconnecting peripheral nodes - called \emph{accessibility (2)}; and (c) interconnecting central nodes - referred as \emph{accessibility (3)}.

\begin{figure*}[!htbp]
	\includegraphics[width=\linewidth]{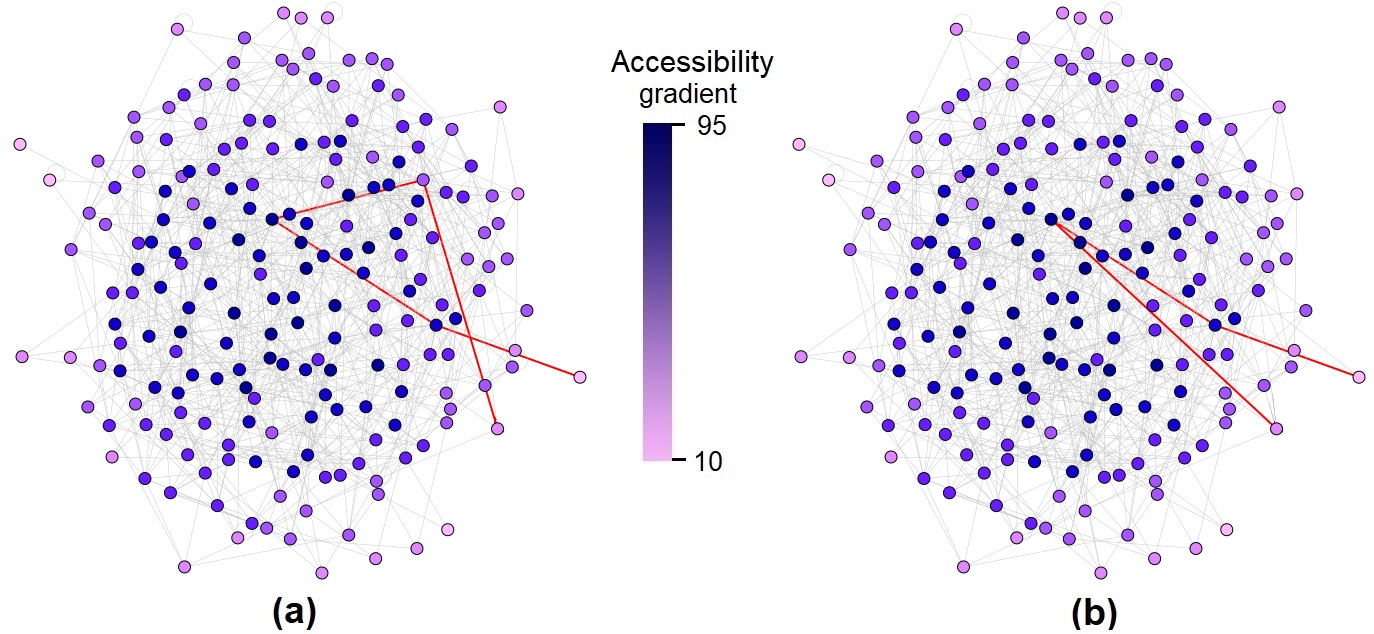}
	\caption{Example of a network, with nodes colored according to their accessibility values, before (a) and after (b) the application of the \emph{accessibility (1)} strategy. The shortest path between a same pair of nodes are indicated in red in both networks.}
		\label{RedesAcess}
\end{figure*}

\section{Results and Discussion}
    
\subsection{Network models}
    
A total of $30$ instances of each model was considered and $50$ iterations of each of our strategies were applied, with the average shortest path recalculated in every step. The variation of the average shortest path length in terms of the iterations is shown in Figure~\ref{resultadoComparacao} for the four adopted topologies.

In every considered case, the accessibility (1), the betweenness centrality and the degree-based strategies yielded better results when compared to the others strategies, being overcome only by the accessibility (3) strategy in the case of the WS models. Also, the WS model was the only case in which the best strategies defined straight curves for the ASPL reduction. For every adopted model, the worst reduction in average shortest path length was observed for the \emph{regular topology} strategy, which involves making the topology more regular.

The degree-based strategy reduces the ASPL by connecting nodes with small degrees to large-degree nodes, the latter being more likely to be part of shorter paths. 

As mentioned, the accessibility can be used to define the concept of a center and a periphery of a network~\cite{travenccolo2008accessibility,travenccolo2009border}. Therefore, as nodes at the network border tend to be more poorly connected between themselves, the shortest path between two peripheral nodes tends to cross the central region of the network, i.e.~the shortest path between two nodes of low accessibility tends to go through nodes of high accessibility, as illustrated in Figure~\ref{RedesAcess}(a). The \emph{accessibility (1)} strategy yielded the best performance for the considered simulations because this strategy, by connecting high and low accessibility nodes, tends to provide shortcuts between peripheral nodes (see Figure~\ref{RedesAcess}(b)).  

The betweenness centrality, allowing the second best performance among the considered strategies, is known to be large for hubs of the networks~\cite{ALBERT}. Hubs represent central nodes that connect most nodes from its surrounds with hubs from others regions of the network, acting as bridges between those regions. Therefore, the betweenness centrality strategy likely acted by enhancing hub connections, facilitating the flow of information between regions, with good impact on the shortest paths of the network.   

When comparing the improvements yielded by the degree-based and the betweenness centrality strategies, which were similar, it follows that, in practice, it could be more interesting to adopt the degree-based approach because of its substantially smaller computational cost, compared to that implied by calculating the betweenness centrality.

Table~\ref{table1} shows the decrease in ASPL relative to its initial value, that is, before the application of the optimization strategies, for each considered model and strategy. Generally speaking, the largest improvements were observed for the WS model. The BA model yielded the smallest improvements.

\begin{table*}[t]
    \centering
    \begin{tabular}{|c|c|c|c|c|c|}
        \cline{3-6}
        \multicolumn{2}{ c| }{}& BA & ER &
        \hspace{4pt} WS & \hspace{4pt} Wax \\
        \hline
        Regular topology & Initial Value & \hspace{1pt} $3.49 \pm 0.03$ \hspace{1pt} & \hspace{1pt} $4.04 \pm 0.04$ \hspace{1pt} & $4.41 \pm 0.02$ & $4.03 \pm 0.04$\\
        & \hspace{1pt} Variation (\%) \hspace{1pt} & $-0.16 \pm 0.02$ & $-0.43 \pm 0.02$ & $-0.60 \pm 0.02$ \hspace{1pt} & $-0.42 \pm 0.02$ \\
        \hline
        Degree & Initial Value & \hspace{1pt} $3.49 \pm 0.03$ \hspace{1pt} & \hspace{1pt} $4.04 \pm 0.04$ \hspace{1pt} & $4.41 \pm 0.02$ & $4.03 \pm 0.04$\\
        & \hspace{1pt} Variation (\%) \hspace{1pt} & $-0.78 \pm 0.05$ & $-1.4 \pm 0.1$ & $-1.38 \pm 0.05$ \hspace{1pt} & $-1.48 \pm 0.08$ \\
        \hline
        Pref.Attachment & Initial Value & $3.49 \pm 0.03$ & $4.04 \pm 0.04$ & $4.41 \pm 0.02$ & $4.03 \pm 0.04$\\  
        & Variation (\%) & $-0.40 \pm 0.05$ & $-0.83 \pm 0.04$ & $-1.31 \pm 0.05$ & $-0.81 \pm 0.05$ \\
        \hline
        Btw. Centrality & Initial Value & $3.49 \pm 0.03$ & $4.04 \pm 0.04$ & $4.41 \pm 0.02$ & $4.03 \pm 0.04$ \\ 
        & Variation (\%) & $-0.86 \pm 0.06$ & $-1.37 \pm 0.08$ & $-1.51 \pm 0.06$ & $-1.51 \pm 0.08$ \\
        \hline
        Accessibility (1) & Initial Value & $3.49 \pm 0.03$ & $4.04 \pm 0.04$ & $4.41 \pm 0.02$ & $4.03 \pm 0.04$ \\ 
        & Variation (\%) & \cellcolor{gray!25}$-1.20 \pm 0.07$ & \cellcolor{gray!25}$-1.5 \pm 0.1$\hspace{2pt} & $-1.52 \pm 0.05$ & \cellcolor{gray!25} $-1.7 \pm 0.1$ \\
        \hline
        Accessibility (2) & Initial Value & $3.49 \pm 0.03$ & $4.04 \pm 0.04$ & $4.41 \pm 0.02$ & $4.03 \pm 0.04$ \\
        & Variation (\%) & $-0.30 \pm 0.02$ & $-0.64 \pm 0.04$ & $-1.06 \pm 0.06$ & $-0.64 \pm 0.04$ \\
        \hline
        Accessibility (3) & Initial Value & $3.49 \pm 0.03$ & $4.04 \pm 0.04$ & $4.41 \pm 0.02$ & $4.03 \pm 0.04$ \\
        & Variation (\%) & $-0.34 \pm 0.08$ & $-0.97 \pm 0.08$ & \cellcolor{gray!25}$-1.7 \pm 0.1$ & $-0.89 \pm 0.09$ \\
        \hline
    \end{tabular}
    \caption{Comparison of the ASPL evolution after $50$ iterations for each strategy. The initial value is calculated before the application of any strategy and the percentage variation of the average shortest path length is related with respect to the initial value. Also shown is the corresponding standard deviation of $30$ realizations for each strategy and model. The largest variation for each model is highlighted with a gray background.}
        \label{table1}
\end{table*}

\begin{figure*}[!htbp]
	\includegraphics[width=\linewidth]{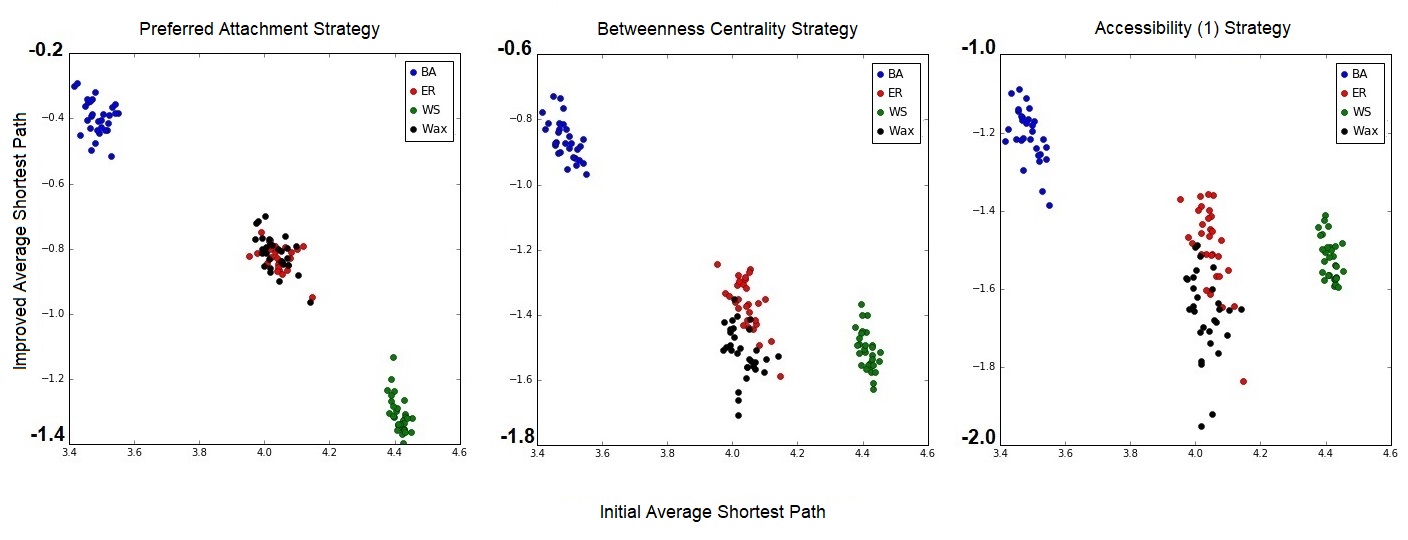}
	\caption{Relationship between the initial average shortest path length found in each of the considered models and the improvement produced by the strategies. Here, only three representative strategies are shown. Note that the improvement axes have distinct ranges among the plots.}
		\label{Correlacao}
\end{figure*}

In order to better understand the relationship between the initial value of the average shortest path length and the relative variation of this property after the optimization strategies, we plot these two quantities as a scatter-plot, which is shown in Figure~\ref{Correlacao}. Generally, the distribution of points in these scatter-plots yields a relatively high correlation between the initial and the improved values for the network models, indicating that a high average shortest path has greater potential to be proportionally reduced. The improvement in ASPL for the ER and WAX models tends to vary more than in the other models, specially when considering the accessibility (1) strategy. 

\subsection{A case study: world-wide airport network}

\begin{figure}[!htbp]
	\includegraphics[width=\linewidth]{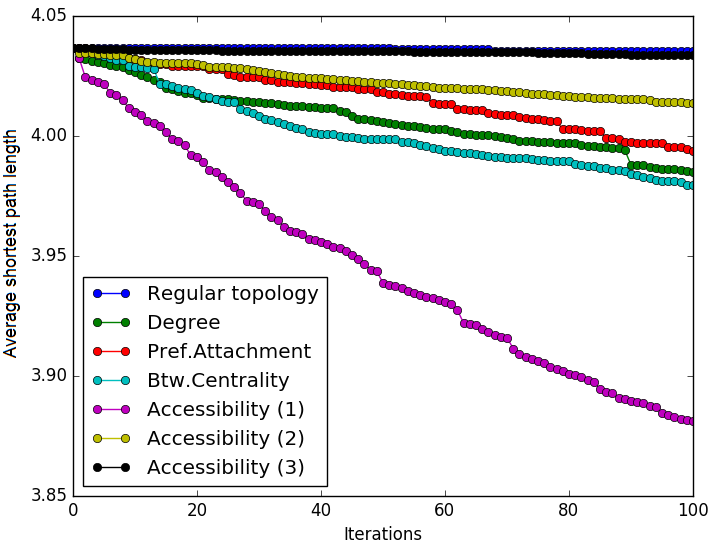}
	\caption{Average shortest path length as a function of added connections obtained by applying the several considered strategies on an airport network.}
		\label{airport}
\end{figure}

In order to study how the application of the proposed strategies can affect real-world systems, we considered a world-wide airport network~\cite{OpenFlights} that represents several airports, which are connected when there are direct flights between them. The results of the application of the strategies can be seen in Figure~\ref{airport}.

The order of strategies performances was almost the same as that found for the network models. The \emph{Accessibility (1)} strategy presented a reduction of $2.43\%$ after $50$ iterations and $3.86\%$ after $100$ iterations, which is considerably larger than what had been obtained for the theoretical models, showing a greater potential for optimization of the average shortest path in that real-world network.

\section{Conclusions}
    
The average shortest path length (ASPL) is an important property of complex networks, as it tends to directly influence the dynamics taking place on a network. For instance, information exchange in a social network tends to be faster when the ASPL is small.~\cite{kwak2010} Thus, it is interesting not only to identify strategies able to efficiently optimize this property, but also to verify the effectiveness of optimization strategies among different networks. 

In this work, we considered $7$ ASPL optimization strategies, all involving the subsequent addition of edges to the network, but using different criteria for connecting pairs of nodes. The degree, betweenness and accessibility values of the nodes were taken into account for placing the new connections. The strategies were applied to the Erd\H{o}s-R\'enyi, Barab\'asi-Albert, Watts-Strogatz and Waxman network models and also to an airport network. The results indicate that, among the considered models, connecting pairs of nodes having large and small accessibility has the best potential for decreasing the ASPL of a network. This happens because nodes having low accessibility tend to be associated with the periphery, or border,~\cite{travenccolo2008accessibility,travenccolo2009border} of the network. Such nodes tend to be poorly interconnected, thus having their communication to dependent on paths passing through more central nodes, characterized by large accessibility. When new connections are added between low and large-accessibility nodes (accessibility (1) strategy), the shortest path distances among the nodes in the periphery rapidly decrease, when compared to other strategies. 

When considering the optimization strategies for different network topologies, the BA model showed large resilience to ASPL reduction. This is likely because networks generated by the BA model already have very low ASPL~\cite{barabasi2000}. The ASPL of the airport network showed relatively small variation, with the striking exception of the accessibility (1) strategy, which sharply decreased the ASPL of this network.

Future works could address the optimization of the ASPL of spatial network models as well as diverse real-world networks, including a cost for adding long range connections to the network. Another possibility would be to verify the influence of the average degree on the optimization strategies, as this measurement is known to have a strong effect on the shortest path distribution.

    \acknowledgements    
G. S. Domingues thanks CNPQ (Grant No. 131909/2019-3) for financial support. Cesar H. Comin thanks FAPESP (grant no. 18/09125-4) for financial support. Luciano da F. Costa thanks CNPq (grant no. 307085/2018-0) and NAP-PRP-USP for sponsorship. This work has been supported also by the FAPESP grant 2015/22308-2.

	\bibliographystyle{apa}
	\bibliography{referencias}
\end{document}